\documentclass[pra,twocolumn,floats,floatfix, showpacs,preprintnumbers,amsmath,amssymb,aps,titlepage]{revtex4}

\usepackage{graphicx,graphics,color,epsfig}
\usepackage{bm}
\usepackage{amsmath}
\usepackage{amssymb}
\newcommand{\bq}{\begin{equation}}
\newcommand{\eq}{\end{equation}}
\newcommand{\bn}{\begin{eqnarray}}
\newcommand{\en}{\end{eqnarray}}

\begin{document}
\title{Generation of pure continuous-variable entangled cluster states of four separate atomic ensembles in a ring cavity}
\author{Gao-xiang \surname{Li}$^{a}$}
\email{gaox@phy.ccnu.edu.cn}
\author{Sha-sha \surname{Ke}$^{a}$}
\author{Zbigniew \surname{Ficek}$^{b}$}
\affiliation{$^{a}$Department of Physics, Huazhong Normal University, Wuhan 430079, China\\
$^{b}$Department of Physics, School of Physical Sciences, The University of Queensland,
Brisbane, Australia 4072}

\begin{abstract}
A practical scheme is proposed for creation of continuous variable
entangled cluster states  of four distinct atomic ensembles located
inside a high-finesse ring cavity. The scheme does not require a set
of  external input squeezed fields, a network of beam splitters and
measurements. It is based on nothing else than the dispersive
interaction between the atomic ensembles and the cavity mode and a
sequential application of laser pulses of a suitably adjusted
amplitudes and phases. We show that the sequential laser pulses
drive the atomic "field modes" into pure squeezed vacuum states. The
state is then examined against the requirement to belong to the
class of cluster states. We illustrate the method on three examples
of the entangled cluster states, the so-called continuous variable
linear, square and T-type cluster states.
\end{abstract}


\maketitle

\section{Introduction}

The investigation of the continuous variable (CV) quantum
information has attracted a great interest along with the
development of different techniques for generation, manipulation and
detection of CV multipartite entangled state~\cite{s1}. The CV
entangled states have many applications in various quantum
information processing such as quantum teleportation~\cite{s2},
dense coding~\cite{s3}, entanglement swapping~\cite{s4}, and quantum
telecloning~\cite{s5}. Recently, Zhang and Braunstein~\cite{s6} have
introduced a kind of~CV Gaussian multipartite entangled states, the
so-called~CV cluster states. The states possess the property that
their entanglement is harder to destroy than that of the CV
Greenberger-Horne-Zeilinger (GHZ) state. Menicucci {\it et
al.}~\cite{s7} have proposed a generalization of the universal
quantum computation by employing the~CV cluster state and an optical
implementation involving squeezed-light sources, linear optics, and
performing a measurement by the homodyne detection. Since then, the
problem of the generation of a CV cluster state has became a very
important issue.  Several proposals have been put
forward~\cite{s8,s9,s10,s11}. Essentially all these proposals are
based on linear optics schemes and measurements on the system. For
example, Su {\it et al.}~\cite{s10} experimentally produced the CV
quadripartite cluster state of the electromagnetic field by
utilizing two amplitude-quadrature and two phase-quadrature squeezed
states of light and linearly optical transformation. Three different
kinds of four-mode CV cluster states, which are suitable for
small-scale implementations of one-way quantum computation, have
been proposed and experimentally constructed by applying squeezed
light sources and a network of beam splitters~\cite{s11}.

Apart from the schemes based on linear optics, a numerous interest
is now in the study of practical application of atomic systems to CV
quantum information and computation~\cite{s12,s13,s14}. This is
because a large collection of identical atoms, called an {\it atomic
ensemble}, can be efficiently coupled to quantum light if a
collective superposition state of many atoms can be utilized for the
coupling~\cite{dcz}. The existence of long atomic ground-state
coherence lifetimes has been thus used to realize the long-lived,
high-fidelity storage of quantum states that is important for
long-distance quantum networking~\cite{s15}. Relying on optical
joint measurements of the atomic ensemble states and magnetic
feedback reconstruction, a protocol to achieve high fidelity quantum
state teleportation of a macroscopic atomic ensemble using a pair of
quantum-correlated atomic ensembles has also been proposed by Dantan
{\it et al.}~\cite{s16}. The dissipation version of the
Lipkin-Meshkov-Glick (LMG) model~\cite{s17} and  the effective Dicke
model~\cite{s18} have been proposed in atomic ensemble based on
cavity-mediated Raman transitions. The quantum trajectories of
collective atomic spin states of driven two-level atoms and coupled
to a cavity field has been studied~\cite{s19}. Especially, it has
been proposed that the unconditional preparation of a two-mode
squeezed state of effective bosonic modes realized in a pair of
atomic ensembles interacting collectively with a two-mode optical
cavity and laser fields~\cite{s20}. However, how to prepare CV
cluster states in atomic ensembles still remains an open question.

In this paper, we address this question and propose a practical
scheme for creation of four-mode CV entangled cluster states of four
separate atomic ensembles located inside a single-mode ring cavity.
The scheme does not involve any external sources of squeezed light
and networks of beam splitters used in the linear optics
schemes~\cite{s11}. In our scheme the atomic ensembles are driven by
laser pulses and a cluster state is created by specific sequential
choices of the Rabi frequencies and phases of the laser pulses. It
is assumed that the atoms interact with the laser pulses and the
cavity field in a highly nonresonant dispersive manner. The
dispersive interaction involves virtual states, not the excited
states of the atoms. Therefore,  the process of preparation of a
cluster state is not affected by the atomic spontaneous emission. We
illustrate the method for three types of cluster states,  the
so-called CV linear, square, and T-type cluster states~\cite{s11}
and demonstrate how these states can be deterministically prepared
using the sequence of laser pulses and the cavity dissipation. In
Sec.~\ref{sec2},  we introduce the model and derive the effective
Hamiltonian of the system. In Sec.~\ref{sec3}, the wave functions of
the three kinds of the four-mode CV cluster states are explicitly
introduced and the concrete steps of preparation of these
quadripartite cluster states are given. We summarize our results in
Sec.~\ref{sec4}.

\section{The Model }\label{sec2}

\begin{figure}[hbp]
\includegraphics[width=\columnwidth,keepaspectratio,clip]{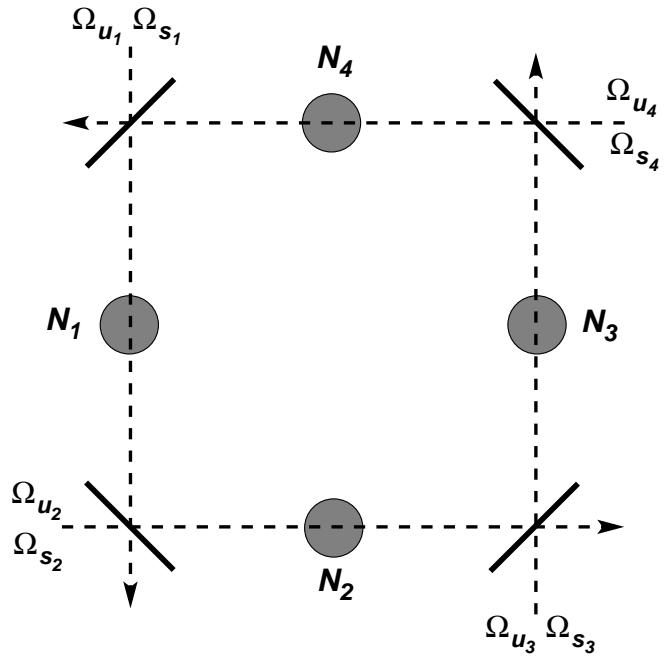}
\caption{Configuration of a ring cavity and four atomic ensembles for the preparation of entangled CV cluster states.  External pulse lasers of the Rabi frequencies $\Omega_{u_j}$ and $\Omega_{s_j}$ couple to only one (propagating) mode of the cavity and interact dispersively with the atoms.}
\label{fig1}
\end{figure}

We consider a system consisting of four atomic ensembles located
inside a high-finesse ring cavity. The cavity is composed of four
mirrors that create two modes, called propagating and
counter-propagating modes, to which the atomic ensembles are equally
coupled. External pulse lasers that are used to drive the atomic
ensembles couple to only a single propagating mode, as it is
illustrated in Fig.~\ref{fig1}. We assume that the atoms are
homogeneously distributed inside the ensembles. In this case only
the forward scattering occurs that allows us to neglect the coupling
of the atoms to the counter-propagating mode and work in the single
mode approximation~\cite{s18,s20}. In other words, the cavity mode propagates with
the laser fields. This condition can be easily fulfilled in trapped room-temperature
atomic ensembles where fast atomic oscillations over the interaction time
lead to a collectively enhanced coupling of the atoms to a single mode that is practically
collinear with the laser fields~\cite{dcz}.

The cavity is damped with the rate $\kappa$ that
is assumed small to achieve a high finesse at a relatively large
size of the cavity. In the current experiments with ring
cavities~\cite{kr03,na03,kl06}, finesses up to $F= 1.7\times 10^{5}$
are achieved with the cavity length (cavity round trip) $L\approx
100$\ mm, which gives the decay rate $(\kappa/2\pi) \approx 20$\
kHz. With the relatively large length of the cavity, the mode
spacing of the cavity field $\Delta\omega \approx 1$\ GHz that is
much larger than the width of the cavity modes. Thus, the one mode
approximation, assumed here, appears to be practical.

The atomic ensembles contain a large number of identical four-level
atoms each composed of two stable ground states,
$|0_{jj^\prime}\rangle$, $|1_{jj^\prime}\rangle$, and two excited
states $|u_{jj^\prime}\rangle$, $|s_{jj^\prime}\rangle$. Here, the
subscript $j \ (j=1,2,3,4)$ labels the atomic ensembles, and the
subscript $j^{\prime} \ (j^\prime=1,2,\ldots,N_j)$ labels individual
atoms in a given ensemble. Such a scheme might be realized in
practice, e.g., by employing alkali atoms, with
$|0_{jj^\prime}\rangle$ and $|1_{jj^\prime}\rangle$ as different
ground-state sublevels.
\begin{figure}[hbp]
\includegraphics[width=\columnwidth,keepaspectratio,clip]{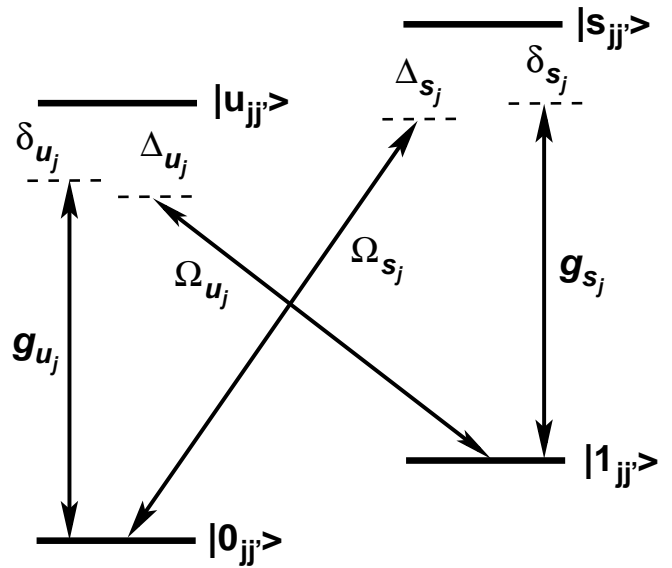}
\caption{Atomic level scheme. Two highly detuned laser fields of the Rabi frequencies $\Omega_{u_j}$ and $\Omega_{s_j}$ drive the atomic transitions $|1_{jj^{\prime}}\rangle\rightarrow |u_{jj^{\prime}}\rangle$ and $|0_{jj^{\prime}}\rangle\rightarrow |s_{jj^{\prime}}\rangle$, respectively. The atomic transitions $|1_{jj^{\prime}}\rangle\rightarrow |s_{jj^{\prime}}\rangle$ and $|0_{jj^{\prime}}\rangle\rightarrow |u_{jj^{\prime}}\rangle$ are coupled to the single-mode cavity with the coupling strengths $g_{u_j}$ and $g_{s_j}$, respectively.}
\label{fig2}
\end{figure}
The ground state $|0_{jj^\prime}\rangle$ of energy
$E_{0_{jj^{\prime}}}=0$  is coupled to the excited state
$|s_{jj^\prime}\rangle$ by a laser field of the Rabi frequency
$\Omega_{s_{j}}$ and frequency $\omega_{Ls_{j}}$ that is detuned
from the atomic transition frequency by $\Delta_{s_j} =
(\omega_{s_j}-\omega_{Ls_j})$, where
$\omega_{s_j}=E_{s_{jj^{\prime}}}/\hbar$ and $E_{s_{jj^{\prime}}}$
is the energy of the state $|s_{jj^\prime}\rangle$. Similarly, the
ground state $|1_{jj^\prime}\rangle$ of energy
$E_{1_{jj^{\prime}}}=\hbar\omega_{1_j}$ is coupled to the excited
state $|u_{jj^\prime}\rangle$ of energy
$E_{u_{jj^{\prime}}}=\hbar\omega_{u_j}$ by an another laser field
with the Rabi frequency $\Omega_{u_j}$, and the angular frequency
$\omega_{Lu_j}$ that is detuned from the atomic transition
$|1_{jj^\prime}\rangle\rightarrow|u_{jj^\prime}\rangle$ by
$\Delta_{u_{j}}= (\omega_{u_j}-\omega_{1_j}-\omega_{Lu_j})$. The
atoms interact with the cavity mode of frequency $\omega_a$ that
simultaneously couples to the
$|u_{jj^\prime}\rangle\leftrightarrow|0_{jj^\prime}\rangle$ and
$|s_{jj^\prime}\rangle\leftrightarrow|1_{jj^\prime}\rangle$
transitions, with the coupling strengths  $g_{u_j}$ and $g_{s_j}$,
respectively. We assume that the atom-field coupling strengths are
uniform through the atomic ensembles. This is consistent with
current experiments involving ring cavities and and large samples of
trapped atoms~\cite{s18}. Practical sizes of the atomic samples are
$\sim 10^{3}$\ nm that is smaller than the cavity mode radius
$w_{0}=130\mu$m, and also is much smaller than the practical length
of a single arm of the cavity.

The system is described by the Hamiltonian, which in the interaction
picture takes the form
\begin{eqnarray}
H_I &=& \sum_{j=1}^4\sum_{j^\prime=1}^{N_j}\left\{\frac{1}{2}\Omega_{u_j}e^{i(\phi_{u_j}+\Delta_{u_j}t)}|u_{jj^\prime}\rangle\langle1_{jj^\prime}|\right. \nonumber\\
&&\left. +\frac{1}{2}\Omega_{s_j}e^{i(\phi_{s_j}+\Delta_{s_j}t)}|s_{jj^\prime}\rangle\langle0_{jj^\prime}|\right. \nonumber \\
&&\left. +g_{u_j}e^{i\delta_{u_j}t}a|u_{jj^\prime}\rangle\langle0_{jj^\prime}|\right. \nonumber \\
&&\left. +g_{s_j}e^{i\delta_{s_j}t}a|s_{jj^\prime}\rangle\langle1_{jj^\prime}|+{\rm H.c.}\right\} ,
\end{eqnarray}
where $\delta_{u_j} = (\omega_{u_j}-\omega_{a})$ and $\delta_{s_j} = (\omega_{s_j}-\omega_{1_j}-\omega_{a})$ are the detunings of the cavity field from the atomic resonances, and $\phi_{s_j}$, $\phi_{u_j}$ are the phases of the driving fields.

In order to eliminate spontaneous scattering of photons to modes
other than the privileged cavity mode, we assume that the laser
fields and the cavity mode frequencies are highly detuned from the
atomic resonances, i.e., the detuning $\{|\Delta_{u_j}|,
|\Delta_{s_j}|, |\delta_{u_j}|, |\delta_{s_j}|\}\gg\{g_{u_j},
g_{s_j}, |\Omega_{u_j}|, |\Omega_{s_j}| \}$. This allows us to apply
the adiabatic approximation~\cite{s17,s18,s19,s20} under which we
eliminate the excited states of the atoms and obtain an effective
Hamiltonian of the atom-field interaction of the form~\cite{s18,s20}
\begin{eqnarray}
H_{{\rm eff}} &=& \sum_{j=1}^4\left\{\frac{\Omega_{u_j}^2}{4\Delta_{u_j}}\left(\frac{N_j}{2}+J_{z_j}\right)
+\frac{\Omega_{s_j}^2}{4\Delta_{s_j}}\left(\frac{N_j}{2}-J_{z_j}\right)\right. \nonumber\\
&+&\left. \!\left[\frac{g_{u_j}^2}{\Delta_{u_j}}\left(\frac{N_j}{2}-J_{z_j}\right)\!+\!\frac{g_{s_j}^2}{\Delta_{s_j}}
\left(\frac{N_j}{2}\!+\!J_{z_j}\!\right)\!+\!\delta_a\right]\!a^\dag a\right. \nonumber \\
&+&\left. \left[a^\dag\left(\beta_{u_j}J_j^-+\beta_{s_j}J_j^\dag\right)+{\rm H.c.}\right]\right\} ,\label{e2}
\end{eqnarray}
where
\begin{eqnarray}
J_j^\dagger &=& \sum_{{j^\prime}=1}^{N_j}|1_{jj^\prime}\rangle\langle0_{jj^\prime}| ,\quad
J_j^{-} = \sum_{{j^\prime}=1}^{N_j}|0_{jj^\prime}\rangle\langle1_{jj^\prime}| ,\nonumber\\
J_{z_j} &=& \frac{1}{2}\sum_{{j^\prime}=1}^{N_j}(|1_{jj^\prime}\rangle\langle1_{jj^\prime}|-|0_{jj^\prime}\rangle\langle0_{jj^\prime}|) ,
\end{eqnarray}
are the collective operators of the atomic ensembles,
\begin{eqnarray}
\beta_{u_j} = \frac{\Omega_{u_j}g_{u_j}}{2\Delta_{u_j}}e^{i\phi_{u_j}} ,\quad
\beta_{s_j}=\frac{\Omega_{s_j}g_{s_j}}{2\Delta_{s_j}}e^{i\phi_{s_j}}
\end{eqnarray}
are the effective coupling constants of the ensembles to the cavity
mode, and $\delta_a=\omega_a-(\omega_{L_{s_j}}+\omega_{L_{u_j}})/2$
is the detuning of the cavity field from the average frequency of
the laser fields.  Here the frequencies of the driving lasers are
also assumed to satisfy $\omega_{Ls_j}-\omega_{Lu_j}=2\omega_{1_j}$.

The first line of Eq.~(\ref{e2}) represents the free energy of the
ensembles. The second line represents an intensity dependent (Stark)
shift of the atomic energy levels, and the third line represents the
interaction between the cavity field and the atomic ensembles. The
essential feature of the effective Hamiltonian (\ref{e2}) is that
the atoms interact dispersively with the cavity mode. This means
that the cavity mode will remain unpopulated during the evolution.
Note the presence of nonlinear terms, $a^{\dagger}J_{j}^{\dagger},\
aJ_{j}^{-}$, analogous to the counter-rotating terms, that will play
the crucial role in creation of a squeezed state between the atomic
ensembles.

The Hamiltonian (\ref{e2}) is general in the parameter values. In
what follows, we will work with a simplified version of the
Hamiltonian by assuming equal coupling constants of the atomic
ensembles to the cavity field, $g_{u_j}=g_{s_j}=g$, and equal
detunings $\Delta_{u_j}=\Delta_{s_j}=\Delta$. We further assume that
each atomic ensemble contains the same number of atoms, i.e.,
$N_j=N$, and we choose the detuning~$\delta_{a}$ such that the
resonant condition with the shifted resonances, $\delta_{a} +
4g^2N/\Delta = 0$, is satisfied.

A standard way to obtain a CV cluster state is to entangle a number
of field (bosonic) modes. Therefore, we shall work in the field
(bosonic) representation of the atomic operators and consider a
procedure to entangle the corresponding number of the bosonic modes.
In this approach, we make use of the Holstein-Primakoff
representation~\cite{s21} that transforms the collective atomic
operators into harmonic oscillator annihilation and creation
operators $c_j$ and $c_{j}^{\dagger}$ of a single bosonic mode
\begin{eqnarray}
J_j^\dagger = c_j^\dag\sqrt{N -c_j^\dag c_j} ,\quad J_{z_j}=c_j^\dag c_j-N/2 .\label{e6}
\end{eqnarray}
For the procedure considered here of the preparation of the cluster
states, the mean number of atoms transferred to the states
$|1_{jj^\prime}\rangle$ is expected to be much smaller than the
total number of atoms in each ensemble, i.e., $\langle c_j^\dag
c_j\rangle \ll N$. By expanding the square root in Eq.~(\ref{e6})
and neglecting terms of the order of $\emph{O}(1/N)$, the collective
atomic operators can be approximated as~\cite{s20}
\begin{eqnarray}
J_j^\dagger\approx \sqrt{N}c_j^\dagger ,\quad J_{z_j}\approx -N/2 .
\end{eqnarray}
Substituting these expressions into Eq.~(\ref{e2}) and omitting the
constant energy terms, we find that the effective Hamiltonian can be
simplified to the form
\begin{eqnarray}
H_{{\rm eff}} &=&  \frac{\sqrt{N}g}{2\Delta}\sum_{j=1}^{4}\left\{ \Omega_{u_j}\left(e^{i\phi_{u_j}}a^\dag c_j+e^{-i\phi_{u_j}}c^{\dag}_{j}a\right)\right. \nonumber\\
&&\left. +\Omega_{s_j}\left(e^{-i\phi_{s_j}}ac_j+e^{i\phi_{s_j}} c^\dag_ja^\dag\right)\right\} .\label{e7}
\end{eqnarray}

We now consider the evolution of the system under the effective
Hamiltonian (\ref{e7}) including also a possible loss of photons due
to the damping of the cavity mode. If the cavity mode is allowed to
decay with a rate $\kappa$, the system then is determined by the
density operator $\rho$ whose the time evolution satisfies the
master equation
\begin{equation}
\dot{\rho}=-i[H_{{\rm eff}} ,\rho]+L_a\rho ,\label{e8}
\end{equation}
where
\begin{equation}
L_a\rho = \frac{1}{2}\kappa(2a\rho a^\dag-a^\dag a\rho-\rho a^\dag a) ,
\end{equation}
represents the damping of the field by the cavity decay~$\kappa$.
Choosing $\Omega_{u(s)_j}$ and $\phi_{u(s)_j}$ appropriately allows
for the preparation of the system in a desired state that then
decays to its steady-state with the rate $\kappa$.

Note that we have ignored the damping of the atoms by spontaneous
emission due to large detunings assumed in the derivation of the
effective Hamiltonian~(\ref{e7}). The spontaneous emission rate due
to off-resonant excitation is estimated at the rate $\gamma_{{\rm
eff}}=\frac{1}{4}(\gamma/2\pi)(\Omega_{r(s)_{j}}/\Delta)^{2}$ that
with the experimental values of $\gamma = 6$\ MHz for a rubidium
atom and $\Omega_{r(s)_{j}}/\Delta = 0.005$ gives $\gamma_{{\rm
eff}} \approx 40$\ Hz. The estimated value for $\gamma_{{\rm eff}}$
is significantly smaller than~$\kappa$ predicted for a cavity of the
finesse $F= 1.7\times 10^{5}$.

The master equation (\ref{e8}) will be use to analyze how to deterministically prepare cluster states for
the four atomic ensembles by a proper choosing of the Rabi frequencies $\Omega_{u(s)_j}$, and the phases~$\phi_{u(s)_j}$. We will use different sequences of the laser pulses to drive the atomic ensembles and assume that all the atoms are initially in the lowest energy state $|0_{jj^\prime}\rangle$.

\section{Generation of the CV quadripartite cluster states}\label{sec3}

\begin{figure}[hbp]
\includegraphics[width=\columnwidth,keepaspectratio,clip]{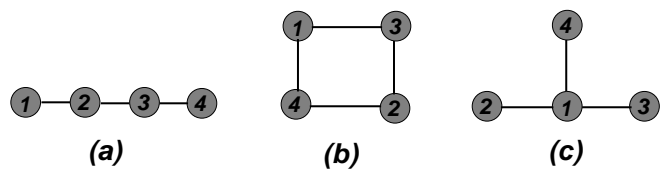}
\caption{Examples of CV quadripartite cluster states; (a) linear cluster state, (b) square cluster state,
and (c) T-shape cluster state.}
\label{fig3}
\end{figure}

We now proceed to discuss the detailed procedure of the preparation of CV quadripartite cluster states of atomic ensembles located inside a single-mode ring cavity. In particular, we shall show how to deterministically prepare a linear cluster state, a square cluster state, and a T-shape cluster state, as shown in
Fig.~\ref{fig3}.

As a preface, we briefly discuss how one could distinguish that a
given state  belongs to the class of cluster states. Simply, a given
state is quantified as a cluster state if the quadrature
correlations are such that in the limit of infinite squeezing, the
state becomes zero eigenstate of a set of quadrature
combinations~\cite{s11}
\begin{eqnarray}
\left(\hat{p}_{a} -\sum_{b\in N_{a}}\hat{x}_{b}\right) \rightarrow 0 ,\label{e10}
\end{eqnarray}
where $\hat{x}$ and $\hat{p}$ are the position and momentum
operators (quadratures) of a mode $a$, and the modes $b$ are the
nearest neighbors $N_{a}$ of the mode $a$. In what follows, we
quantify a given state as a cluster state by evaluating the
variances of linear combinations of the momentum and position
operators of the involved field modes. If the variances vanish in
the limit of the infinite squeezing then, according to the above
definition, a given state is a cluster state.

\subsection{The preparation of the linear CV cluster state}

Let us first concentrate on the preparation of a four-mode linear
cluster state involving four separate atomic ensembles, as shown in
Fig.~\ref{fig3}(a). As we shall see, the preparation process depends
crucially on the specific choice of the Rabi frequencies and phases
of the driving laser fields.

In order to unconditionally prepare a linear cluster state involving
four separate atomic ensembles and find its explicit form, we make
an unitary  transformation $d_{L_j}=T_1c_jT_1^\dag$ $(j=1,2,3,4)$
that transfers the $c_j$ operators into linear combinations
\begin{eqnarray}
d_{L_1} &=& -\frac{1}{\sqrt{2}}(ic_1+c_2) ,\nonumber\\
d_{L_2} &=& -\frac{1}{\sqrt{10}}(ic_1-c_2-2ic_3-2c_4) ,\nonumber\\
d_{L_3} &=& -\frac{1}{\sqrt{2}}(c_3+ic_4) ,\nonumber\\
d_{L_4} &=& -\frac{1}{\sqrt{10}}(2c_1+2ic_2+c_3-ic_4) .
\end{eqnarray}
Since the operators $d_{L_j}$ commutate with each other, the
combined modes are orthogonal to each other. This will allow us to
prepare each mode separately in a desired state. In other words, an
arbitrary transformation performed on the operators of a given mode
will not affect the remaining modes.

Let us now illustrate the procedure of preparing the field modes in
a desired state by using laser pulses of equal length and suitably
chosen magnitudes of the Rabi frequencies and phases of the driving
lasers. More concretely, for a cluster state, all the modes
$d_{L_{j}}$ should be prepared in a squeezed vacuum state. We employ
the fact that the modes can be separately prepared in the squeezed
vacuum state. Afterwards, if the variances of specific combinations
of the quadrature components of the field operators vanish in the
limit of the infinite squeezing, the desired state is a linear CV
cluster state.

In the first step of the preparation, we send a set of laser pulses
driving the the atomic ensembles 1 and 2 only, the lasers driving
the atomic ensembles 3 and 4 are turned off. We choose the Rabi
frequencies as
\begin{eqnarray}
\Omega_{u_n} &=& \frac{\Omega_{s_n}}{r}=\sqrt{2}\Omega ,\quad n=1,2 , \nonumber\\
\Omega_{u_j} &=& \Omega_{s_j}=0 ,\quad j=3,4 ,
\end{eqnarray}
and the phases of the driving lasers
\begin{eqnarray}
\phi_{u_1} &=& \frac{3}{2}\pi ,\quad \phi_{s_1}=\frac{1}{2}\pi ,\nonumber \\
\phi_{u_2} &=& \phi_{s_2}=\pi .
\end{eqnarray}

After the first step of the preparation, the mode $d_{L_{1}}$ is
left in a state that is described by the density operator $ \rho
_1=T_1\rho T_1^\dag$, which obeys a master equation
\begin{equation}
\frac{d}{dt} \rho _1 = -i\beta\left[\left(a^\dagger d_{L_1}+ra^\dagger
d_{L_1}^\dagger\right)+{\rm H.c.}, \rho _1\right] + L_a \rho _1 ,\label{e16}
\end{equation}
where $\beta = \sqrt{N}g\Omega/\Delta$ and $r\in (0,1)$. If we now
perform the single-mode squeezing transformation for the mode
$d_{L_1}$ as $\tilde{\rho}_1=S^\dag_1(\xi)\rho_1S_1(\xi)$ with the
single-mode squeezing operator\cite{fds}
\begin{eqnarray}
S_j(\xi)\!=\!\exp\!\left[\frac{\xi}{2}\left( d_{L_j}^2-d_{L_j}^{\dag2}\right)\right] ,
\end{eqnarray}
where $\xi = \tanh^{-1}(r)$, we find that the master equation~(\ref{e16}) becomes
\begin{eqnarray}
\frac{d}{dt}\tilde{\rho}_1=\beta\sqrt{1-r^2}\left[ad_{L_1}^\dag+a^\dag
d_{L_1},\tilde{\rho}_1\right]+L_a\tilde{\rho_1} .\label{e18}
\end{eqnarray}
It can be shown from Eq.~(\ref{e18}) that in the steady state, the
cavity mode will be in the vacuum state and the mode $d_{L_1}$ will
be in a squeezed vacuum state. This is easy to understand, the
master equation (\ref{e18}) represents two coupled modes with the
cavity mode linearly damped with the rate $\kappa$. Since the
remaining modes $d_{L_2}$, $d_{L_3}$, and $d_{L_4}$ are decoupled
from the mode $d_{L_1}$, they remain in an undetermined state
$\rho_{d_{L_2}d_{L_3}d_{L_4}}(\tau)$.

In order to estimate the required time to reach the steady state,
we calculate eigenvalues of Eq.~(\ref{e18})
\begin{eqnarray}
\lambda_\pm =
-\frac{\kappa}{2}\pm\left[\left(\frac{\kappa}{2}\right)^2-\beta^2(1-r^2)\right]^\frac{1}{2}
\end{eqnarray}
from which we observe that as long as $\beta\sqrt{1-r^2}>\kappa/2$,
the time for the system  to reach its steady state is of order of
$\sim 2/\kappa$\cite{s22}. Therefore, the system will definitely
evolve into the steady state, provided the interaction time is
sufficient long, for example, $\tau=4/\kappa$. The time
$\tau=4/\kappa$ determines the time scale in our protocol for the
preparation of the cluster states.

By taking the inverse unitary transformation, it follows that in the
steady state, the total system is in a state determined by the
density operator
\begin{eqnarray}
\rho_1\!(\tau)\!=\!S_1(\xi)|0_a,\!0_{d_{L_1}}\!\rangle
\langle0_a,\!0_{d_{L_1}}\!|S_1^\dag(\xi)\!\otimes\!\rho_{d_{L_2}\!d_{L_3}\!d_{L_4}}\!(\tau) .
\end{eqnarray}
Briefly summarize what we have obtained after the first step of the
preparation of a linear CV cluster state. The application of
suitably chosen laser pulses leaves the mode $d_{L_1}$ prepared in
the single-mode squeezed vacuum state, with the cavity field found
in the vacuum state, and the remaining combined modes $d_{L_2}$,
$d_{L_3}$, and $d_{L_4}$ left in the states related to their initial
states.

In the second step, we turn off the first series of driving lasers
and sequentially send another series of laser pulses with different
parameters to preparation of the combined bosonic mode $d_{L_2}$ in
the single-mode squeezed vacuum state so that a similar linearly
mixing interaction between the cavity mode and another combined
bosonic modes arises. As before, for the first series of pulses, a
single-mode squeezed vacuum state for this combined bosonic mode can
be prepared due to the cavity dissipation. The Rabi frequencies of
the second series of the laser pulses, which are turned on during
the time of $t\in[\tau,2\tau)$ for the preparation of the combined
bosonic mode $d_{L_2}$ in the single-mode squeezed vacuum state
$S_2(\xi)|0_{d_{L_2}}\rangle$, are
\begin{eqnarray}
\Omega_{u_n} &=& \frac{\Omega_{s_n}}{r}=\frac{2}{\sqrt{10}}\Omega ,\quad n = 1,2 , \nonumber\\
\Omega_{u_j} &=& \frac{\Omega_{s_j}}{r}=\frac{4}{\sqrt{10}}\Omega ,\quad j = 3,4 ,
\end{eqnarray}
and the phases of the driving lasers
\begin{eqnarray}
\phi_{u_1} &=& \frac{3}{2}\pi ,\quad \phi_{s_1}=\frac{1}{2}\pi ,\nonumber\\
\phi_{u_3} &=& \frac{1}{2}\pi ,\quad \phi_{s_3}=\frac{3}{2}\pi ,\nonumber\\
\phi_{u_2} &=& \phi_{s_2}=\phi_{u_4}=\phi_{s_4} = 0 .
\end{eqnarray}
The specific choice of the Rabi frequencies and the phases ensures
the mode $d_{L_2}$ to be prepared in the state described by the
density operator $\rho_{1}$ satisfying the same master as
Eq.~(\ref{e16}) but with $d_{L_1}$ replaced by $d_{L_2}$.

In the third step, which is performed during the time of
$t\in[2\tau,3\tau)$, the combined bosonic mode $d_{L_3}$ is being
prepared in the single-mode squeezed vacuum state
$S_3(\xi)|0_{d_{L_3}}\rangle$. The Rabi frequencies of the third
series of the laser pulses, are
\begin{eqnarray}
\Omega_{u_n} &=& \frac{\Omega_{s_n}}{r} = 0 ,\quad n=1,2 , \nonumber\\
\Omega_{u_j} &=& \frac{\Omega_{s_j}}{r} = \sqrt{2}\Omega ,\quad j=3,4 ,
\end{eqnarray}
and the phases are
\begin{eqnarray}
\phi_{u_1} &=& \phi_{s_1} = \phi_{u_2} = \phi_{s_2} = 0,\nonumber\\
\phi_{u_3} &=& \frac{3}{2}\pi ,\quad \phi_{s_3}=\frac{1}{2}\pi ,\nonumber \\
\phi_{u_4} &=& \phi_{s_4} = \pi .
\end{eqnarray}

The fourth series, laser pulses are turned on during the time of
$t\in[3\tau,4\tau)$. In this final series, the combined bosonic mode
$d_{L_4}$ is prepared in the single-mode squeezed vacuum state
$S_4(\xi)|0_{d_{L_4}}\rangle$. The laser pulses required to achieve
this are of the Rabi frequencies
\begin{eqnarray}
\Omega_{u_n} &=& \frac{\Omega_{s_n}}{r}=\frac{4}{\sqrt{10}}\Omega ,\quad n=1,2 ,\nonumber\\
\Omega_{u_j} &=& \frac{\Omega_{s_j}}{r}=\frac{2}{\sqrt{10}}\Omega ,\quad j=3,4 ,
\end{eqnarray}
and phases
\begin{eqnarray}
\phi_{u_k} &=& \phi_{s_k}=0 ,\quad k=1,3 ,\nonumber \\
\phi_{u_2} &=& \frac{1}{2}\pi ,\quad \phi_{s_2}=\frac{3}{2}\pi ,\nonumber \\
\phi_{u_4} &=& \frac{3}{2}\pi ,\quad \phi_{s_4} = \frac{1}{2}\pi .
\end{eqnarray}
Therefore after enough long time $4\tau$, the system evolves into the state determined by the density operator
\begin{equation}
\rho_1(4\tau)= |\Phi_L\rangle\langle \Phi_L| \otimes
 |0_a\rangle\langle0_a| ,
\end{equation}
where
\begin{eqnarray}
|\Phi_{L}\rangle &=& T_1|\Psi_L\rangle =
S_1(\xi)|0_{d_{L_1}}\rangle
\otimes S_2(\xi)|0_{d_{L_2}}\rangle \nonumber \\
&& \otimes  S_3(\xi)|0_{d_{L_3}}\rangle
\otimes  S_4(\xi)|0_{d_{L_4}}\rangle
\end{eqnarray}

Evidently, the transformed state is in the form of a quadripartite squeezed vacuum state~\cite{fds}.
Once all the combined modes $d_{L_j}$ are prepared in the corresponding single-mode
squeezed vacuum states, the explicit form of the state created can be obtained by reversing the unitary transformation $T_1$. This leads to a pure linear~CV quadripartite state of the form
\begin{eqnarray}
|\psi_L\rangle&=&\exp\left\{-\frac{\xi}{10}[c_1^2-c_2^2-c_3^2+c_4^2
-8i(c_{1}c_2+c_3c_4)\right. \nonumber\\
&&\left. -4(c_1+ic_2)(c_3-ic_4)]-{\rm H.c.}\} |\{0_{c}\right\}\rangle, \label{e13}
\end{eqnarray}
where
\begin{eqnarray}
 |\{0_{c}\}\rangle = |0_{c_{1}},0_{c_{2}},0_{c_{3}},0_{c_{4}}\rangle .
\end{eqnarray}

We may summarize that by an appropriate driving the four atomic
ensembles, the four  combined bosonic modes will be ultimately
prepared in four single-mode squeezed vacuum states. It can be done
in four steps by appropriately choosing the laser parameters such as
phases and intensities. Then, by applying an inverse unitary
transformation, the pure linear CV entangled state $|\Psi_L\rangle$
for the four atomic ensembles can be obtained.

The question remains as to whether the state $|\psi_L\rangle$ is an
example of the linear CV cluster state. To examine this, we
introduce the quadrature amplitude $q_j=(c_j+c_j^\dag)/\sqrt{2}$ and
phase $p_j = -i(c_j-c_j^\dag)/\sqrt{2}$ components of the four modes
involved, and easily find that the variances of the linear
combinations of the components, evaluated according to the
definition (\ref{e10}), are
\begin{eqnarray}
V(p_1-q_2)&=&e^{-2\xi},\nonumber\\
V(p_2-q_1-q_3)&=&\frac{3}{2}e^{-2\xi},\nonumber\\
V(p_3-q_2-q_4)&=&\frac{3}{2}e^{-2\xi},\nonumber\\
V(p_4-q_3)&=&e^{-2\xi} ,
\end{eqnarray}
where $V(X)=\langle X^2\rangle-\langle X\rangle^2$. Clearly, all the
four variances tend to zero in the limit of infinite squeezing,
$\xi\rightarrow\infty$. We therefore conclude that the state
$|\Psi_L\rangle$ is a four-mode linear CV cluster state~\cite{s11}.

\subsection{The preparation of the square CV cluster state}

Having discussed the procedure of creation of a linear cluster
state, we now proceed to describe the realization of an entangled CV
square cluster state $|\psi_S\rangle$. The strategy is based on the
same procedure used above for the preparation of the CV linear
cluster state.

A square CV cluster state may be obtained by performing a unitary
transformation $d_{S_j}=T_2c_jT_2^\dagger$ to obtain superposition
modes determined by the following operators
\begin{eqnarray}
d_{S_1}&=&-\frac{1}{\sqrt{10}}(ic_1+ic_2+2c_3+2c_4),\nonumber\\
d_{S_2}&=&-\frac{i}{\sqrt{2}}(c_1-c_2),\nonumber\\
d_{S_3}&=&-\frac{1}{\sqrt{10}}(2c_1+2c_2+ic_3+ic_4),\nonumber\\
d_{S_4}&=&-\frac{i}{\sqrt{2}}(c_3-c_4).
\end{eqnarray}

Similar to the preparation of the CV linear cluster state for the
four separated atomic ensembles, here we can make the interaction
between the cavity and the four mode $c_j$ $(j=1,2,3,4)$ reduce to
the coupling between the mode $a$ and one of the combined mode
$d_{S_j}$ in the form of a combined linear mixing process by
choosing proper parameters of lasers. After a long-time interaction
$\tau=4/\kappa$, the cavity mode $a$ and the combined mode $d_{S_j}$
will respectively develop into the vacuum state and the squeezed
vacuum  state $\exp(\frac{\xi}{2} d_{S_j}^2-{\rm
H.c.})|0_{d_{S_j}}\rangle$ due to the cavity dissipation. Therefore
through separately sending four series of driving lasers and send
another series of laser pulses with different appropriate parameters
such as phases and amplitudes, the four single-mode squeezed vacuum
states for the combined bosonic modes $d_{S_j}$ can be prepared with
the help of the cavity dissipation. Subsequently, by reserving the
above unitary transformations, we can deterministically prepare the
quadripartite square cluster state $|\psi_S\rangle$.

The four sets of the laser pluses are chosen as follows. In the time
period of $t\in[0,\tau)$, we apply the first set of pulses of the parameters
\begin{eqnarray}
\Omega_{u_n} &=& \frac{\Omega_{s_n}}{r}=\frac{2}{\sqrt{10}}\Omega ,\quad n = 1,2 , \nonumber\\
\Omega_{u_j} &=& \Omega_{s_j}=\frac{4}{\sqrt{10}}\Omega ,\quad j = 3,4 ,
\end{eqnarray}
and the phases of the driving lasers
\begin{eqnarray}
\phi_{u_n} &=& \frac{3}{2}\pi ,\quad \phi_{s_n}=\frac{1}{2}\pi ,\quad  n=1,2 ,\nonumber\\
\phi_{u_j} &=& \phi_{s_j}=\pi ,\quad j=3,4 .
\end{eqnarray}

At the time of $t=\tau$, we turn off the first set of lasers and send the second set of laser
pluses of the Rabi frequencies
\begin{eqnarray}
\Omega_{u_n} &=& \frac{\Omega_{s_n}}{r}=\sqrt{2}\Omega ,\quad n = 1,2 , \nonumber\\
\Omega_{u_j} &=& \Omega_{s_j}=0 ,\quad j = 3,4 ,
\end{eqnarray}
and the phases
\begin{eqnarray}
\phi_{u_1}\!=\! \frac{3}{2}\pi ,\quad \phi_{s_1}=\frac{1}{2}\pi ,\quad
\phi_{u_2} = \frac{1}{2}\pi ,\quad \phi_{s_2}=\frac{3}{2}\pi .
\end{eqnarray}

At the time of $t=2\tau$, we turn off the second set of laser pulses and
send the third set of the Rabi frequencies
\begin{eqnarray}
\Omega_{u_n} &=& \frac{\Omega_{s_n}}{r}=\frac{4}{\sqrt{10}}\Omega,\quad n = 1,2 , \nonumber\\
\Omega_{u_j} &=& \Omega_{s_j}=\frac{2}{\sqrt{10}}\Omega ,\quad j = 3,4 ,
\end{eqnarray}
and the phases
\begin{eqnarray}
\phi_{u_1} &=& \phi_{u_2} = \phi_{u_3}=\phi_{u_4}=\frac{3}{2}\pi ,\nonumber \\
\phi_{s_1} &=& \phi_{s_2}=\pi ,\quad \phi_{s_3} = \phi_{s_4}=\frac{1}{2}\pi .
\end{eqnarray}

At the time $t=3\tau$, we turn off the third set of lasers and send the fourth set of laser
pulses of the Rabi frequencies
\begin{eqnarray}
\Omega_{u_n} &=& \frac{\Omega_{s_n}}{r}= 0 ,\quad n = 1,2 , \nonumber\\
\Omega_{u_j} &=& \Omega_{s_j}= \sqrt{2}\Omega ,\quad j = 3,4 ,
\end{eqnarray}
and the phases
\begin{eqnarray}
\phi_{u_n}\!=\! \phi_{s_n} &=& 0 ,\quad n=1,2 ,\nonumber \\
\phi_{u_3} = \frac{3}{2}\pi ,\quad \phi_{s_3} &=& \!\frac{1}{2}\pi ,\quad \!\phi_{u_4}\!=\!\frac{1}{2}\pi ,\quad
\!\phi_{s_4}\!=\!\frac{3}{2}\pi .
\end{eqnarray}

After the above sequence of the laser pulses, the system is left in
a state described by the density operator
\begin{equation}
\rho_1(4\tau)=T_2|\Psi_S\rangle\langle \Psi_S|T_2^\dag \otimes
 |0_a\rangle\langle0_a|.
\end{equation}
Inverting the transformation $T_2$, we find that the system is in the state
\begin{eqnarray}
|\psi_S\rangle &=& \exp\left\{-\frac{\xi}{10}[c_1^2+c_2^2+c_3^2+c_4^2-8c_1c_2-8c_3c_4\right. \nonumber\\
&&\left. -4i(c_1+c_2)(c_3+c_4)]-{\rm H.c.}\right\}|\{0_{c_j}\}\rangle.
\end{eqnarray}

Finally, we calculate the variances in the sum and difference
operators  following the general procedure for quantifying the
cluster states and find that the variances for the square type state
illustrated in Fig.~\ref{fig2}(b) are given~by
\begin{eqnarray}
&& V(p_1-q_3-q_4) = V(p_2-q_3-q_4) = V(p_3-q_1-q_2)\nonumber \\
&& = V(p_4-q_1-q_2) = \frac{3}{2}e^{-2\xi} .
\end{eqnarray}
We see that the variances tend to zero when the squeezing parameter
goes to infinity, $\xi\rightarrow\infty$. Hence the state
$|\Psi_S\rangle$ is an example of a four-mode square cluster
state~\cite{s11}. We conclude that the sequential application of the
laser pulses, the CV entangled quadripartite square cluster state
$|\psi_S\rangle$ is unconditionally produced.

\subsection{The preparation of a T-shape cluster state}

Finally, we apply the above formalism to describe how to prepare  a
T-shape entangled cluster state of the form
\begin{eqnarray}
|\psi_T\rangle&=&\exp\left\{\frac{\xi}{4}[c_1^2-c_2^2-c_3^2-c_4^2+2ic_1(c_2+c_3+c_4)\right. \nonumber\\
&&\left. +2(c_2c_3+c_2c_4+c_3c_4)]-{\rm H.c.}\}|\{0_{c_j}\right\}\rangle .\label{e40}
\end{eqnarray}

As above, we first perform a unitary transformation
$d_{T_j}=T_3c_jT_3^\dagger$ that transfers the bosonic operators
$c_{j}$ into superposition modes
\begin{eqnarray}
d_{T_1}&=&\frac{\sqrt{3}}{2}\left[ic_1-\frac{1}{3}(c_2+c_3+c_4)\right],\nonumber\\
d_{T_2}&=&\frac{\sqrt{6}}{3}\left[c_2-\frac{1}{2}(c_3+c_4)\right],\nonumber\\
d_{T_3}&=&\frac{\sqrt{2}}{2}\left(c_3-c_4\right),\nonumber\\
d_{T_4}&=&\frac{1}{2}\left(ic_1+c_2+c_3+c_4\right).
\end{eqnarray}
We will show that the transformation $T_3$ leads to the field  modes
that can be prepared in a T-shape cluster state.

Following the similar four-step procedure as in the above two examples,
we first show that by a suitable choice of the Rabi frequencies and
phases of the laser fields, one can prepare the modes $d_{T_j}$ in a
squeezed vacuum state.

To achieve this, we choose the following sequence of the laser
pulses. In the first step, the atomic ensemble is driven by pulse
lasers over the  time period of $t\in[0,\tau)$, with the Rabi
frequencies
\begin{eqnarray}
\Omega_{u_1} &=& \frac{\Omega_{s_1}}{r}= 0 , \nonumber\\
\Omega_{u_n} &=& \frac{\Omega_{s_n}}{r} =\frac{\sqrt{3}}{3}\Omega ,\quad n = 2,3,4 ,
\end{eqnarray}
and the phases
\begin{eqnarray}
\phi_{u_1} &=& \frac{1}{2}\pi ,\quad \phi_{s_1}=\frac{3}{2}\pi ,\nonumber\\
\phi_{u_n} &=& \phi_{s_n}=\pi ,\quad j= 2,3,4 .
\end{eqnarray}

At the time of $t=\tau$, we turn off the first set of lasers and
send a second set of the pulses, with the Rabi frequencies
\begin{eqnarray}
\Omega_{u_1} &=& \Omega_{s_1} = 0 , \nonumber\\
\Omega_{u_2} &=& \Omega_{s_2}= \frac{2\sqrt{6}}{3}\Omega ,\nonumber \\
\Omega_{r_n} &=& \frac{\Omega_{s_n}}{r}=\frac{\sqrt{6}}{3}\Omega ,\quad j=3,4 ,
\end{eqnarray}
and the phases
\begin{eqnarray}
\phi_{u_1}\!=\! \frac{3}{2}\pi ,\quad \phi_{s_1}=\frac{1}{2}\pi ,\quad
\phi_{u_2} = \frac{1}{2}\pi ,\quad \phi_{s_2}=\frac{3}{2}\pi .
\end{eqnarray}
The following sets of pulses are: At the time of $t=2\tau$:
\begin{eqnarray}
\Omega_{u_n} &=& \frac{\Omega_{s_n}}{r}= 0 ,\quad n=1,2 , \nonumber\\
\Omega_{u_j} &=& \frac{\Omega_{s_j}}{r} =\sqrt{2}\Omega ,\quad j = 3,4 ,\nonumber\\
\phi_{u_1} &=&  \phi_{s_1}= \phi_{u_2} =\phi_{s_2} =0 ,\nonumber\\
\phi_{u_j} &=& \phi_{s_j}=\pi ,\quad j=  3,4 .
\end{eqnarray}
At the time $t=3\tau$
\begin{eqnarray}
\Omega_{u_n} &=& \frac{\Omega_{s_n}}{r}= \Omega ,\quad n=1,2,3,4 , \nonumber\\
\phi_{u_1} &=&  \frac{1}{2}\pi ,\quad \phi_{s_1}=\frac{3}{2}\pi  ,\nonumber\\
\phi_{u_j} &=& \phi_{s_j}= 0 ,\quad j=  2,3,4 .
\end{eqnarray}
After the above sequence of the laser pulses, the system is left in
a state described by the density operator
\begin{equation}
\rho_1(4\tau)=T_3|\Psi_T\rangle\langle \Psi_T|T_3^\dag \otimes
 |0_a\rangle\langle0_a|.
\end{equation}
Inverting the transformation $T_3$, we find that the system is in the pure state (\ref{e40}).

The only thing left is to determine as to whether the state
$|\Psi_T\rangle$ belongs  to the class of cluster states. It is done
by calculating the variances of the sum and difference operators
from which we find that for the state $|\psi_T\rangle$, the
variances are given by
\begin{eqnarray}
V(p_1-q_2-q_3-q_4) &=& 2e^{-2\xi},\nonumber\\
V(p_2-q_1)\! =\!V(p_3-q_1) &=& V(p_4-q_1)\!=\!e^{-2\xi} .
\end{eqnarray}
Clearly, the variances tend to zero when $\xi\rightarrow\infty$. Consequently, we may conclude that
the state $|\Psi_T\rangle$ is an example of a four-mode continuous variable T-shape cluster state.

\section{Conclusions}\label{sec4}

We have described a practical scheme for the preparation of
entangled CV cluster states of effective bosonic modes realized in
four physically separated atomic ensembles interacting collectively
with a single-mode optical cavity and driving laser fields. We have demonstrated 
robustness of the scheme on three examples of the so-called continuous variable 
linear, square and T-type cluster states.
The basic idea of the scheme is to transfer the ensemble field modes into suitable
linear combinations that can be prepared, by a sequential
application of the laser pulses, in pure squeezed vacuum states. We
have shown, by referring to practical parameters, that the scheme is
feasible with the current experiments.

\section*{Acknowledgements}
This work is supported by the National Natural Science Foundation of
China (under Grant Nos 10674052 and 60878004), the Ministry of
Education under project NCET (under Grant No NCET-06-0671) and SRFDP
(under grant no. 200805110002), and the National Basic Research
Project of China (2005CB724508).

\end{document}